\documentclass[prb]{revtex4}%
\usepackage{graphicx}
\usepackage{amsmath}
\usepackage{amsfonts}
\usepackage{amssymb}%
\providecommand{\U}[1]{\protect\rule{.1in}{.1in}}
\providecommand{\U}[1]{\protect\rule{.1in}{.1in}}
\begin{document}
\title{On the statistical physics of metastable and non-equilibrium stationary states}
\author{Alejandro Cabo$^{1,2}$ and Sergio Curilef$^{3}$}
\affiliation{$^{1}$\textit{Group of Theoretical Physics, Instituto de Cibern\'{e}tica,
Matem\'{a}tica y F\'{\i}sica, La Habana, Cuba}}
\affiliation{$^{2}$ \textit{International Centre for Theoretical Physics, Strada Costiera
11, Miramare, Trieste, Italy }}
\affiliation{$^{3}$\textit{Facultad de F\'{\i}sica, Universidad Cat\'{o}lica del Norte,
Antofagasta, Chile }}

\begin{abstract}
\noindent The optimization problems defining meta-stable or
stationary equilibrium are explored. The Gibbs scheme is modified
aiming to describe the statistical properties of a class of
non-equilibrium and metastable states. The system is assumed to
maximize the usual definition of the Entropy, subject to the
standard constant energy and norm restrictions, plus additional
constraints. The central assumption is that the existence of the
considered metastable state is determined by the action of \ this
additional dynamical constraint, that blocks the evolution of the
system up to its maximum \ Entropy state, thus maintaining it in
the metastable or stationary configuration. After requiring from
the statistical description\ to be valid for the combination of
two nearly independent subsystems, it follows that the eigenvalues
of the constraint operators $C$ should have the restricted
homogeneous form $C(p_{i})=p_{i}^{q}$, in terms of the eigenvalues
of the density matrix $p_{i}$, where $q$ is a fixed real number.
Therefore, the distribution of the eigenvalues of the constraint
function have the Tsallis structure. This conclusion suggests the
interpretation of the $q$ parameter as reflecting the homogeneous
dependence of the constraint determining the metastable state on
the density matrix. An application to the plasma experiments of
Huang and Driscoll is expected to be considered elsewhere in order
to compare the results with Tsallis scheme and the minimal
Enstrophy one.

\bigskip


\ \bigskip

\end{abstract}
\maketitle



\section{Introduction}

The statistical physics of systems which remains for long times in metastable
and stationary states is an important theme of research nowadays
[\onlinecite{boghosian,driscoll,plastinos}]. Since the appearance of the
proposals for variation of the Boltzman statics proposed by Tsallis for a wide
class of non-equilibrium states, it has a been an explosion of interest in the
field [\onlinecite{tsallis}]. A particular point of attention in the
literature is devoted to the understanding of the meaning of the special real
parameter $q$ which fully characterize the deviation of this new statistical
description from the Boltzman one [\onlinecite{qo,q1}].

In the present work we explore the statistical properties of metastable and
non-equilibrium stationary states with the aim of determining possible new
ways of studying these properties and their possible connections with the
conceptual elements in the Tsallis approach. Out central starting assumption
is that statistical properties differing from those of the Gibbs thermal
equilibrium states, could be generated, at least in a large class of systems,
by the existence of a dynamically generated constraint  stopping for a while
their relaxation to the thermal equilibrium, described by the Bloch density
matrix.\ \ Since the systems under consideration are in quasi-equilibrium in
their metastable states, it is here assumed that the additional constraint is
approximately conserved. \ Then, the procedure being proposed here, consists
in assuring that the system on its evolution, tends to maximize the usual
Entropy in terms of the density matrix $S=-k$ $Tr[\rho\log\rho], $ under the
standard restrictions of constant energy and norm, plus an additional
specially constructed constraint.

The commutation of the statistical density matrix $\rho$ describing the
metastable state and the Hamiltonian is assumed since the physical quantities
of the metastable state are quasi time independent.\ Consider now the
evolution of the system in the quasi-equilibrium state. It is clear that the
probability distribution of the energy values of the Gibbs subsystems will not
follow the one given by the Bloch density matrix. There will be a different
distribution along the metastable evolution defined by a density matrix $\rho
$. This behavior is assumed here to be produced by the existence of a
constraint $C$ which constant value during the movement means its
conservation. That is, it will be assumed to commute with the Hamiltonian.
Also considering that the energy spectrum is non-degenerated, the three
operators $H,\rho$ and $C$ can be simultaneously diagonalized. This property
allows to express the constraint $C$ as a function of the density matrix
$C[\rho]$. Taking into account that  the product $C[\rho H$ is also a
conserved quantity, a natural way for imposing the effects of the constraint
$C$ in the variational Gibbs problem arises: \ when maximizing the Entropy the
constant value of the trace of $C[\rho] H$ will we also required. In other
words, this form of the new restriction will added in this paper onto the
standard Gibbs maximization problem defining the statistical density matrix by
the conditional extremum point of the Entropy as a functional of the density matrix.

As a consequence of the above definitions, the interesting result follows that
the additivity of the modified conserved mean values, for two approximately
independent subsystems, directly implies that the constraint function
$C[\rho]$ \ should have the Tsallis homogeneous structure $C(\rho)=C_{q}%
\rho^{q}$, with $q$ being an arbitrary real number. The possibility that the
density matrices solving the new conditional extremum problem corresponds to
metastable states is furnished by the fact that, by assumption, the constraint
$C$ is a weakly conserved quantity, which can be unstable because in general,
there could no exist a rigorously conserved dynamical variable associated to
it. However, it could has the chance of locally constraint the system for a
while, to be in a metastable state.

\ The equations of the extremum problem for the determination of the density
matrix describing the considered metastable or non equilibrium stationary
state are also presented.

The work will proceed as follows. In Section 2 the proposal of the scheme is
presented. Section 3 is devoted to argue that, if the statistical description
of physical system satisfies the additivity condition, then the dependence of
the constraint $C$ on the probabilities have the Tsallis structure. The
explicit form of the extremum equations determining the density matrix of the
metastable state are also presented in this Section. Finally, in the summary,
the conclusions are reviewed and some possible extensions of the work commented.

\section{Statistical mechanics of some metastable and stationary states}

Lets us consider the quantum description of a physical systems \ having a
dynamics fixed \ by a Hamiltonian operator $H.$ \ \ \ In the Gibbs approach
the properties of the systems in thermal equilibrium are described by the
Bloch density matrix
\begin{equation}
\rho=\exp(-\frac{H}{kT}),
\end{equation}
satisfying $[H,\rho]=0$ \ and which is the conditional maximum of the Entropy
functional under constant mean energy. That is, the Bloch density matrix is
the extremum of the functional
\begin{align}
S &  =-k\ Tr[\rho\log(\rho)]+\alpha(Tr[\rho H]-E)+\nonumber\\
&  +\beta(Tr[\rho]-1),
\end{align}
in which $\alpha$ and $\beta$ are auxiliary Lagrange multipliers for imposing
the conservation of the energy $E$ and the total probability equal to $1$, and
k is the Boltzman constant.

Let us assume that during a relatively large relaxation time $\tau,$ the
system is not allowed to approach the thermal equilibrium state, because it is
in a metastable or non-equilibrium stationary state. \ We will assume that
this metastable state is \ produced by a dynamically conserved quantity $C$
which approximately restricts the rapid evolution of the system to the Gibbs
thermal equilibrium. \ The approximate satisfaction of this constraint for
large times will be represented here by its, also approximate, commutation
with the Hamiltonian.
\[
\lbrack H,C]=0.
\]

The stationary character of the density matrix $\rho$ will be assumed to also
imply $[H,\rho]=0.$ As mentioned above, in order to simplify the discussion,
let us consider that the spectrum of the Hamiltonian is not degenerate.
\ Therefore, $H,$ $\rho$ and $C $ are \ all diagonalized in the common basis
of eigen-functions of $H$. Henceforth, $C$ can be expressed as a certain
function of the density matrix $C=C[\rho].$

\ But, due to the above definitions the quantities $C[\rho]H$ is also
conserved
\begin{align}
\lbrack C[\rho]H,H]  & =0,\nonumber\\
\lbrack C[\rho],H]  & =0.
\end{align}
Therefore, a convenient way emerges for including the effects of the
constraint $C$ in the Gibbs maximization problem determining an extremum of
the entropy $S$, in the considered metastable or stationary state. The
proposal is to include the constancy of the specially defined quantity \ \
\begin{equation}
\frac{Tr[C(\rho)H]}{Tr[C(\rho)]},
\end{equation}
through a corresponding Lagrange multiplier in the extremum problem. \ \ Then,
adding this new constraint to the expression of the Entropy after to be
multiplied by the corresponding Lagrange multiplier, the conditional Entropy
functional takes the form
\begin{align}
S &  =-kTr[\rho\log(\rho)]+\alpha(Tr[\rho H]-E)+\nonumber\\
&  +\beta(Tr[\rho]-1)+\gamma(\frac{Tr[C(\rho)H]}{Tr[C(\rho)]}-E_{C}%
)\nonumber\\
&  =-k\sum_{i}p_{i}\log(p_{i})+\alpha(\sum_{i}p_{i}\epsilon_{i}-E)+\beta
(\sum_{i}p_{i}-1)+\nonumber\\
&  +\gamma(\frac{\sum_{i}C(p_{i})\epsilon_{i}}{\sum_{i}C(p_{i})}-E_{F}).
\end{align}

In what follows we will assume two possibilities: \

1) \ The $\alpha$ parameter is non vanishing and becomes an effective Lagrange
multiplier. This corresponds to impose both: the mean energy conservation
constraint in common with the $C$ dependent constraint.

2) \ The $\alpha$ parameter is taken as vanishing and then, it is disregarded
as a Lagrange multiplier. This variant corresponds to only fixing the
conservation of the new constraint in addition to the constant probability one.

\section{Tsallis $q$ parameter from additivity of the description}

Let us now \ consider the implications on the modified problem produced by
requiring the additivity of the statistical description. That is, the
condition will be imposed that the application of the scheme to a combination
of two nearly independent systems, each of them being in the same kind of
metastable state, should be equivalent to the separate application to each one
of the systems. \

Therefore, consider two similar and weakly interacting subsystems, each of
them being in a metastable state of the same sort. The Entropy and the
constraint functions of the first system after the density matrix furnishing
the maximum has been found, are given by
\begin{align}
S^{(1)} &  =-k\sum_{i}p_{i}^{(1)}\log(p_{i}^{(1)}),\nonumber\\
E^{(1)} &  =\sum_{i}p_{i}^{(1)}\epsilon_{i}^{(1)},\text{ \ \ \ }1=\sum
_{i}p_{i}^{(1)},\nonumber\\
E_{C}^{(1)} &  =\frac{\sum_{i}C(p_{i}^{(1)})\epsilon_{i}^{(1)}}{\sum
_{i}C(p_{i}^{(1)})]}.
\end{align}

Analogously for the second system, these same quantities take the form
\begin{align}
S^{(2)} &  =-k\ \sum_{i}p_{i}^{(2)}Log(p_{i}^{(2)}),\nonumber\\
E^{(2)} &  =\sum_{i}p_{i}^{(2)}\epsilon_{i}^{(2)},\text{ \ \ \ }1=\sum
_{i}p_{i}^{(2)},\nonumber\\
E_{C}^{(2)} &  =\frac{\sum_{i}C(p_{i}^{(2)})\epsilon_{i}^{(2)}}{\sum
_{i}C(p_{i}^{(2)})]}.
\end{align}
\ The description is assumed to be also valid for the combined system. In this
case the Entropy and constraints for the composite body are
\begin{align}
S^{(1,2)} &  =-k\sum_{(i,j)}p_{(i,j)}^{(1,2)}\log(p_{(i,j)}^{(1,2)}%
),\nonumber\\
E^{(1,2)} &  =\sum_{(i,j)}p_{(i,j)}^{(1,2)}\epsilon_{(i,j)}^{(1,2)},\text{
\ \ \ }1=\sum_{(i,j)}p_{(i,j)}^{(1,2)},\nonumber\\
E_{C}^{(1,2)} &  =\frac{\sum_{(i,j)}C(p_{(i,j)}^{(1,2)})\epsilon
_{(i,j)}^{(1,2)}}{\sum_{(i,j)}C(p_{(i,j)}^{(1,2)})},
\end{align}
where \ $(i,j)$ symbolizes the couple of indices defining the states of the
combined system $|i,j\rangle=|i\rangle\times|j\rangle$. Then, the independence
of the systems allows to write for the probability of the combined state
$|i,j\rangle$ the relations
\begin{align}
p_{(i,j)}^{(1,2)} &  =p_{i}^{(1)}p_{j}^{(2)},\nonumber\\
\sum_{(i,j)}p_{i}^{(1)}p_{j}^{(2)} &  =\sum_{i}p_{i}^{(1)}\sum_{j}p_{j}%
^{(2)}=1.
\end{align}
Therefore, the constant probability constraint of the individual components
imply the same property \ for the combined system.\

\ \ \ The additivity of the Entropies follows in similar way as usual
\begin{align}
S^{(1,2)} &  =-k\sum_{(i,j)}p_{(i,j)}^{(1,2)}\log(p_{(i,j)}^{(1,2)}%
)\nonumber\\
&  =-k\sum_{(i,j)}p_{i}^{(1)}p_{j}^{(2)}\log(p_{i}^{(1)}p_{j}^{(2)}%
)\nonumber\\
&  =-k\sum_{i}p_{i}^{(1)}\log(p_{i}^{(1)})-k\sum_{j}p_{j}^{(2)}\log
(p_{j}^{(2)})\nonumber\\
&  =S^{(1)}+S^{(2)},
\end{align}
as well as the addition of the mean energies
\begin{align*}
E^{(1,2)} &  =\sum_{(i,j)}p_{(i,j)}^{(1,2)}\epsilon_{(i,j)}^{(1,2)},\\
&  =\text{ }\sum_{(i,j)}p_{i}^{(1)}p_{j}^{(2)}(\epsilon_{i}^{(1)}+\epsilon
_{j}^{(2)})\\
&  =\sum_{(i,j)}p_{i}^{(1)}p_{j}^{(2)}\epsilon_{i}^{(1)}+\sum_{(i,j)}%
p_{i}^{(1)}p_{j}^{(2)}\epsilon_{j}^{(2)}\\
&  =\sum_{i}p_{i}^{(1)}\epsilon_{i}^{(1)}+\sum_{j}p_{j}^{(2)}\epsilon
_{j}^{(2)}\\
&  =E^{(1)}+E^{(2)}.
\end{align*}
\ $\ \ $Now, we will also assume that the modified $C$ dependent mean value
also satisfies the statistical independence condition for the combined system
\begin{equation}
C_{(i,j)}^{(1,2)}=C_{i}^{(1)}C_{j}^{(2)}.\label{indep}%
\end{equation}
where $C_{(i,j)}^{(1,2)}=C(p_{(i,j)}^{(1,2)})$, $C_{i}^{(1)}=C(p_{i}^{(1)})$
and $C_{i}^{(2)}=C(p_{i}^{(2)})$. Therefore, the specially defined mean value
also satisfies the additivity properties as follows
\begin{align}
E_{C}^{(1,2)} &  =\frac{\sum_{(i,j)}C_{(i,j)}^{(1,2)}\epsilon_{(i,j)}^{(1,2)}%
}{\sum_{(i,j)}C_{(i,j)}^{(1,2)}},\nonumber\\
&  =\text{ }\frac{\sum_{(i,j)}C_{i}^{(1)}C_{j}^{(2)}(\epsilon_{i}%
^{(1)}+\epsilon_{j}^{(2)})}{\sum_{(i,j)}C_{i}^{(1)}C_{j}^{(2)}}\nonumber\\
&  =\frac{\sum_{(i,j)}C_{i}^{(1)}C_{j}^{(2)}\epsilon_{i}^{(1)}}{\sum
_{(i,j)}C_{i}^{(1)}C_{j}^{(2)}}+\frac{\sum_{(i,j)}C_{i}^{(1)}C_{j}%
^{(2)}\epsilon_{j}^{(2)}}{\sum_{(i,j)}C_{i}^{(1)}C_{j}^{(2)}}\nonumber\\
&  =\frac{\sum_{i}C_{i}^{(1)}\epsilon_{i}^{(1)}}{\sum_{i}C_{i}^{(1)}}%
+\frac{\sum_{j}C_{j}^{(2)}\epsilon_{j}^{(2)}}{\sum_{i}C_{i}^{(1)}}\nonumber\\
&  =E_{C}^{(1)}+E_{C}^{(2)}.
\end{align}
However, this condition also imposes a strong restriction on the possible
forms of the function $C$ defining the modified mean value. In order to see
this, the statistical independence condition (\ref{indep}) can be rewritten in
the form%
\begin{equation}
C(p_{i}^{(1)}p_{j}^{(2)})=C(p_{i}^{(1)})C(p_{j}^{(2)}).\label{product}%
\end{equation}
Considering $C$ as expanded in powers
\begin{equation}
C(x)=x^{\nu}\sum_{n=0}^{\infty}f_{n}\text{ }x^{n}%
\end{equation}
and substituting this relation in (\ref{product}), it follows \ \
\[
(p_{i}^{(1)}p_{j}^{(2)})^{\nu}\sum_{n=0}^{\infty}f_{n}(\text{ }p_{i}%
^{(1)})^{n}(\text{ }p_{j}^{(2)})^{n}=(p_{i}^{(1)})^{\nu}(p_{j}^{(2)})^{\nu
}\sum_{n=0}^{\infty}\sum_{m=0}^{\infty}f_{n}f_{m}(\text{ }p_{i}^{(1)}%
)^{n}(\text{ }p_{i}^{(2)})^{m},
\]
which after cancelling the common factors can be rewritten as
\begin{equation}
0=\sum_{n=0}^{\infty}\sum_{m=0}^{\infty}f_{n}(\text{ }p_{i}^{(1)})^{n}(\text{
}p_{i}^{(2)})^{m}(\delta_{nm}-f_{m}).
\end{equation}
Next , after taking into account the completeness of the basis of powers of
the variables, and assuming that the values of the $p_{i}$ variables form a
continuous for all values of $i,$ it follows \
\begin{equation}
f_{n}(\delta_{nm}-f_{m})=0,\text{ \ \ for all \ }m\text{ and }n\text{ .}%
\end{equation}
Henceforth, assuming that a particular $f_{m_{o}}\neq0,$ implies that for all
$\ \ m\neq m_{o}$ the $f_{m}$ \ expansion parameters are equal to zero. Thus,
the only non-vanishing coefficient is \ the $f_{m_{o}}.$one. Consequently, the
allowed forms of $\ C$ are
\begin{align}
C(\rho) &  =f_{m_{o}}\rho^{m_{o}+\nu}\nonumber\\
&  =f_{q}\text{ }\rho^{q}.
\end{align}

\subsection{Extremum equations}

After the form of the function $C$ has been determined, the Entropy functional
as modified to impose the constraints through the Lagrange multipliers
procedure, can be written in the more explicit form
\begin{align}
S &  =-k\sum_{i}p_{i}\log(p_{i})+\alpha(\sum_{i}p_{i}\epsilon_{i}%
-E)+\beta(\sum_{i}p_{i}-1)+\nonumber\\
&  +\gamma(\frac{\sum_{i}p_{i}^{q}\epsilon_{i}}{\sum_{i}p_{i}^{q}}-E_{q}).
\end{align}
where the index $C$ in \ \ $E_{F}$ has been changed to $q$ $.$

The extremum equations are
\begin{align}
\frac{\partial S}{\partial p_{i}} &  =0,\text{ \ }\frac{\partial S}%
{\partial\alpha}=0,\text{ \ \ }i=1,2...\infty\nonumber\\
\frac{\partial S}{\partial\beta} &  =0,\text{ \ \ }\frac{\partial S}%
{\partial\gamma}=0,
\end{align}
which after their explicit evaluation \ leads to the following set of coupled
equations for the eigenvalues of the density matrix \ $\rho$%

\begin{align}
k-\alpha\text{ }\epsilon_{i}-\beta &  =-k\log(p_{i})+\gamma\frac{q\text{
}p_{i}^{q-1}}{\sum_{i}p_{i}^{q}}(\epsilon_{i}-E_{q}),\nonumber\\
E &  =\sum_{i}p_{i}\epsilon_{i},\nonumber\\
1 &  =\sum_{i}p_{i},\nonumber\\
E_{q} &  =\frac{\sum_{i}p_{i}^{q}\epsilon_{i}}{\sum_{i}p_{i}^{q}}.
\end{align}
Multiplying the \ first equation by $p_{i}$ \ and \ summing over $i$, a
relation between the Entropy and the lagrange multipliers \ follows
\begin{align}
0 &  =-k\ \sum_{i}p_{i}\log(p_{i})+\gamma\sum_{i}\frac{q\text{ }p_{i}^{q}%
}{\sum_{i}p_{i}^{q}}(\epsilon_{i}-E_{q})-1+\alpha E+\beta,\nonumber\\
&  =-k\sum_{i}p_{i}\log(p_{i})-k+\alpha E+\beta\nonumber\\
&  =\ S+\alpha E-k+\beta.
\end{align}
As usual it could be employed for constructing a generalization of the Free
Energy and other Thermodynamical Potentials. However, we will delay this
discussion to future extensions of the work.

\section{Summary}

A statistical description of metastable states is proposed. The main idea is
that a large class of metastable and stationary states could be associated to
the existence of quasi-conserved constraints blocking the approach of the
system to thermal equilibrium. \ Assuming that this is the case, it is shown
that if the statistical properties of the system have the additivity property,
the eigenvalues of the constraint function in terms of the probabilities have
the Tsallis structure \ $p_{i}^{q}.$ \ This outcome indicates the
interpretation of the Tsallis parameter as furnishing the degree of
homogeneity of the constraint as a function of the density matrix. Finally,
the proposed description is defined by the density matrix which maximizes the
usual expression of the Entropy under the conservation of the mean energy,
normalization and the additional constraint assumed to be enforcing the
metastable state.

In future extensions of the work, it is planned to determine the predictions
of the analysis for the plasma experiments of Huang and Driscoll
[\onlinecite{driscoll}]. It is expected that the procedure could appropriately
describe the experimental results of Ref. [\onlinecite{driscoll}]. Those
results were reasonably explained by the minimization of the so-called
Enstrophy function [\onlinecite{driscoll}] and also by a study employing the
Tsallis statistics in Ref. [\onlinecite{boghosian}]. \ Thus, the consideration
of this task seems to be an appropriate step in checking the predictions of
the discussion presented here.

\section{Acknowledgments}

One of the authors would like to acknowledge the Condensed Matter Section of
the Abdus Salam ICTP and its Head Prof. Vladimir Kravtsov, by the kind
invitation to visit the Centre during August 2007. In addition the helpful
support received by one of the authors (A.C.) from the network of the OEA
(ICTP) on: \textit{Quantum Mechanics Particles and Fields} (Net-35), is also
greatly appreciated. The kind hospitality received by both authors
from the Faculty of Physics of the Pontifical Catholic University of Chile
(PUC, Santiago, Chile) during one stay nearly two years ago, in which the
theme of the work was conceived, is also very much recognized. Finally,
 the very helpful remarks of Prof. M. Marsili (ICTP,Trieste) are  gratefully
  acknowledged.


\bigskip

\begin{thebibliography}{9}                                                                                                %
\bibitem {tsallis}C. Tsallis, \textit{Phys. Lett.}\ \textbf{A195, }329
(1994)\textbf{.}

\bibitem {boghosian}B. Boghosian, \textit{Phys. Rev}. \textbf{E 53, }4754
(1996).\textbf{\ }

\bibitem {driscoll}.X.-P. Huang and C. F. Driscoll, \textit{Phys. Rev. Lett
}\textbf{72},2187 (1994).

\bibitem {plastinos}A. R. Plastino and A. Plastino Phys. Lett. A 174 , 384 (1993).

\bibitem {qo}S. Umarov, C. Tsallis, M. Gell-Mann and S. Steinberg,
cond-mat/0606040v1, 1 Jun 2006.

\bibitem {q1}L. G. Moyang, C. Tsallis and M. Gell-Mann, cond-mat/0509229v2, 16
Dec 2005.
\end{thebibliography}
\end{document}